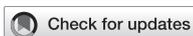

# A tomographic approach to the sum uncertainty relation and quantum entanglement in continuous variable systems

Soumyabrata Paul*, S. Lakshmibala, V. Balakrishnan and
S. Ramanan

Department of Physics and Centre for Quantum Information, Communication and Computing (CQuICC),
Indian Institute of Technology Madras, Chennai, India

Entropic uncertainty relations (EURs) have been examined in various contexts, primarily in qubit systems, including their links with entanglement, as they subsume the Heisenberg uncertainty principle. With their genesis in the Shannon entropy, EURs find applications in quantum information and quantum optics. EURs are state-dependent, and the state has to be reconstructed from tomograms (which are histograms readily available from experiments). This is a challenge when the Hilbert space is large, as in continuous variable (CV) systems and certain hybrid quantum (HQ) systems. A viable alternative approach therefore is to extract as much information as possible about the unknown quantum state directly from appropriate tomograms. Many variants of EURs can be extracted from tomograms, even for CV systems. In earlier work we have defined many tomographic entanglement indicators (TEIs) that can be readily calculated from tomograms without knowledge of the density matrix, and have reported on their efficacy as entanglement indicators in various contexts in both CV and HQ systems. The specific objectives of the present work are as follows: (i) To use the tomographic approach to investigate the links between EURs and TEIs in CV and HQ systems as they evolve in time. (ii) To identify the TEI that most closely tracks the temporal evolution of EURs. We consider two generic systems. The first is a multilevel atom modeled as a nonlinear oscillator interacting with a quantized radiation field. The second is the Λ-atom interacting with two radiation fields. The former model accomodates investigations on the role of the initial state of the field and the ratio of the strengths of interaction and nonlinearity in the connection between TEIs and EURs. The second model opens up the possibility of examining the connection between mixed state bipartite entanglement and EURs, when the number of atomic levels is finite. Since the tomogram respects the requirements of classical probability theory, this effort also sheds light on the extent to which TEIs reflect the temporal behaviour of those EURs which are rooted in the Shannon entropy.

KEYWORDS

entropic uncertainty relations, quantum entanglement, tomographic entanglement indicators, continuous-variable systems, hybrid quantum systems

# 1 Introduction

The inherently probabilistic nature of quantum laws and the manner in which they manifest themselves in the measurement problem are typically formulated in terms of uncertainty relations. The original form of these relations, involving products of variances of incompatible observables (Heisenberg, 1927; Robertson, 1929; Schrodinger, 1930), has been extended in the literature to wider settings, e.g., as in (Maccone and Pati, 2014; Chen and Fei, 2015). More recently the uncertainty relations have been expressed in a manner which subsumes variance-based relations, and formulated in a readily applicable form in quantum information theory. The first hint of such a formulation came through Everett's employment of the Shannon entropy for this purpose (Everett, 1957).

Following this, Hirschman formalized the entropic uncertainty relation (EUR) (Hirschman, 1957). Since then, several EURs have been proposed in the literature [for a detailed review, see Coles et al. (2017)], the version given in (Maassen and Uffink, 1988) being the most well known of these. While the variance-based uncertainty relations can be recovered from EURs, the converse is not possible in general, thus lending more significance to EURs. Further, other generalizations such as the Rényi (Bialynicki-Birula, 2006), Tsallis (Wilk and Wlodarczyk, 2009), and Wehrl (Floerchinger et al., 2021) entropies have also been used to formulate EURs. Uncertainty relations for coarse-grained measurements have been examined in detail in the literature [For a review see, e.g., Toscano et al. (2018)]. Apart from their fundamental importance, EURs find practical use in diverse problems in quantum information such as those pertaining to quantum key distribution, quantum cryptography and entanglement witnesses (Berta et al., 2010).

The link between EURs and quantum entanglement is of immense interest as this is a possible route to understanding correlations and to quantify entanglement (Giovannetti, 2004; Gühne and Lewenstein, 2004; Floerchinger et al., 2021). The experiment on entanglement-assisted entropic uncertainty (Li et al., 2011), carried out on polarization states of entangled photons, has revealed the importance of the nature of the observer in estimating the extent of uncertainty. Experiments on diamond validate results on entropic uncertainty relations for multiple measurements in qubit systems (Xing et al., 2017). Entropy-based and coherence-based uncertainty relations have been tested in an optical platform using Bell-like states and Bell-like diagonal states (Ding et al., 2020). Lower bounds on distillable entanglement that can be measured in viable experiments on cold atoms have been derived in (Bergh and Gärtner, 2021). Experiments to obtain bounds on quantum uncertainty relations have been performed in an all-optical set up with appropriate qubits (Liu et al., 2022), invoking quantum coherence and relative entropy of coherence.

It is to be emphasized, however, that from both theoretical studies and experimental investigations some partial understanding about the interplay between entanglement and EURs, is only available at present for qubit systems. In the context of continuous variable (CV) systems, EURs have been examined theoretically and tight bounds obtained for quadrature observables in optics [see, e.g., (Hertz and Cerf, 2019)]. However, the link between entanglement and EURs in CV systems has not been examined in sufficient detail, even in bipartite cases. This is primarily because of the challenges faced in reconstructing the state from appropriate experimentally obtained tomograms for high dimensional Hilbert spaces. As a consequence, density-operator-based entanglement measures such as the subsystem von Neumann entropy (SVNE) and the subsystem linear entropy (SLE) are not readily calculable. In fact, even for small Hilbert space dimensions, quantum state reconstruction is an arduous task (Hou et al., 2016). The problem is exacerbated if the instantaneous state of a system needs to be obtained in order to examine the changes in the entanglement during time evolution. While state reconstruction procedures in CV systems and in multiple spin arrays are now augmented with machine learning protocols, these attempts are still in their infancy. It is therefore advantageous to avoid detailed state reconstruction wherever possible, and to extract reliable entanglement indicators solely from tomograms. Such entanglement indicators have been proposed in the literature and their efficacy examined in CV and HQ systems by comparing them with standard entanglement measures and monotones (Sharmila et al., 2020; Sharmila et al., 2022).

In this work, we look at CV and hybrid quantum (HQ) systems to examine possible links between EURs and entanglement, relying only on tomograms for this purpose. More specifically, in this paper we examine the efficacy of TEIs in capturing the behaviour of EURs during dynamical evolution of two generic systems. This exercise facilitates identification of the appropriate TEIs that track EURs in the case of both pure and mixed states. We point out that this is the first and essential step in an extended program, that is, expected to shed light on the manner in which several factors affect the interplay between EURs and entanglement *in CV and HQ systems*. These factors include: the number of atomic levels, the extent of nonlinearity and interaction strengths in the relevant Hamiltonian, the degree of coherence of the initial states, bipartite vs. tripartite interactions, and pure vs. mixed states. The importance and novelty of this exercise are enhanced by the fact that even in qubit systems with small Hilbert spaces, the links between EURs and entanglement measures *during dynamics* are rather poorly understood.

The contents of this paper are arranged as follows. In Section 2 we briefly review the salient features of optical tomograms, entanglement indicators, and entropy-based sum uncertainty relations. In Section 3 we consider two generic theoretical models of light-matter interaction, and illustrate the role played by the number of atomic levels, the atom-field interaction and nonlinearities, in establishing the links between uncertainty relations and entanglement. With this aim, in Section 3.1 we investigate a bipartite system of an atom interacting with a radiation field, while Section 3.2 deals with a tripartite model of a Λ-atom interacting with two radiation fields. An optical tomographic indicator capturing the sum entropy dynamics is identified for each of the two systems. In Section 4 we conclude with a brief summary and outlook.

# 2 Formalism: tomograms, entanglement indicators, and uncertainty relations

Consider a single-mode radiation field with photon creation and annihilation operators $\hat{a}^{\dagger}$ and $\hat{a}$ respectively. The set of rotated quadrature operators (Ibort et al., 2009) is defined by

$$\hat{\mathbb{X}}_\theta = \left(\hat{a}^\dagger e^{i\theta} + \hat{a} e^{-i\theta}\right)\Big/\sqrt{2}, \tag{1}$$

where $\theta$ ($0 \le \theta < \pi$) is the phase of the local oscillator in the standard homodyne measurement setup. It is evident that $\theta = 0$ and $\pi/2$ respectively correspond to the $x$ and $p$ quadratures. Equation 1 constitutes a quorum of observables which carry complete information about a given state $\hat{\rho}$. The optical tomogram $w$ ($X_\theta$, $\theta$) (Lvovsky and Raymer, 2009) is given by

$$w(X_\theta, \theta) = \langle X_\theta, \theta | \hat{\rho} | X_\theta, \theta \rangle, \tag{2}$$

where

$$\hat{\mathbb{X}}_\theta | X_\theta, \theta \rangle = X_\theta | X_\theta, \theta \rangle. \tag{3}$$

For each value of $\theta$, the set $\{|X_\theta, \theta\rangle\}$ comprises a complete basis. The optical tomogram visualized with $X_\theta$ as the abscissa and $\theta$ as the ordinate is essentially a collection of histograms corresponding to the quadrature operators. It satisfies the property

$$\int_{-\infty}^{\infty} dX_\theta \ w(X_\theta, \theta) = 1, \ \forall \ \theta. \tag{4}$$

It is advantageous to expand $w$ ($X_\theta$, $\theta$) in terms of the Hermite polynomials for ease in numerical computations (Filippov and Man'ko 2011). For a bipartite system $AB$, the corresponding quadrature operators are defined as

$$\hat{\mathbb{X}}_{\theta_A} = \frac{\left(\hat{a}^\dagger e^{i\theta_A} + \hat{a} e^{-i\theta_A}\right)}{\sqrt{2}}, \quad \hat{\mathbb{X}}_{\theta_B} = \frac{\left(\hat{b}^\dagger e^{i\theta_B} + \hat{b} e^{-i\theta_B}\right)}{\sqrt{2}}, \tag{5}$$

where $(\hat{a}^\dagger, \hat{a})$ (resp. $(\hat{b}^\dagger, \hat{b})$) are the creation and annihilation operators for subsystem $A$ (resp. $B$). The two-mode optical tomogram is given by

$$w_{AB}\left(X_{\theta_A}, \theta_A; X_{\theta_B}, \theta_B\right) = \langle X_{\theta_A}, \theta_A; X_{\theta_B}, \theta_B | \hat{\rho}_{AB} | X_{\theta_A}, \theta_A; X_{\theta_B}, \theta_B \rangle, \tag{6}$$

where $\hat{\rho}_{AB}$ is the two-mode density matrix. The two-mode tomogram satisfies

$$\int_{-\infty}^{\infty} dX_{\theta_A} \int_{-\infty}^{\infty} dX_{\theta_B} \ w_{AB}\left(X_{\theta_A}, \theta_A; X_{\theta_B}, \theta_B\right) = 1, \ \forall \quad \theta_A, \theta_B. \tag{7}$$

The reduced tomogram (corresponding to subsystem $A$, for instance) is given by

$$w_A(X_{\theta_A}, \theta_A) = \int_{-\infty}^{\infty} dX_{\theta_B} \ w_{AB}\left(X_{\theta_A}, \theta_A; X_{\theta_B}, \theta_B\right) \tag{8}$$

for any given $\theta_B$. A similar definition holds good for subsystem $B$. Extension of these definitions to multipartite systems is straightforward.

Both qualitative identification and quantitative estimates of nonclassical effects such as squeezing and entanglement properties of radiation field states can be obtained solely from tomograms. In what follows, we focus on entanglement indicators. While these are not measures, it has been established that they suffice to capture the gross features of bipartite entanglement. We briefly describe, below, two of these "tomographic entanglement indicators" ("TEIs") that we will use in the sequel. An interesting and useful feature of these indicators is that they can be defined for specific tomographic slices (also referred to as "sections"), by choosing appropriate values of $\theta_A$ and $\theta_B$. Averaging over a judiciously chosen set of such indicators provides a section-independent assessment of entanglement. We will exploit this aspect in understanding the connection, if any, between EURs (which are in any case slice-dependent as they relate specific quadrature uncertainties), on the one hand, and both the slice-dependent and averaged entanglement indicator, on the other.

An interesting slice-dependent indicator denoted by $\epsilon_{\text{IPR}}(\theta_A, \theta_B)$ is inspired by the well known inverse participation ratio (IPR) which quantifies the delocalization of a state in a given basis. It was initially proposed to assess the extent of spatial delocalization of atomic vibrations in a specified eigenbasis of a disordered system (Bell and Dean, 1970). Since then, the IPR has been examined in different settings. In particular, in the context of disordered spin chains, general conditions have been derived relating the extent of multipartite entanglement with the IPR computed over a maximal set of mutually unbiased basis states. The details are reported in (Viola and Brown, 2007). Therefore, by its very definition, IPR can be computed in different basis sets. This feature makes it readily computable directly from tomograms. In particular, in CV tomograms, since every value of $\theta$ defines a complete basis set, IPR can be computed in any chosen basis set. This provides a slice-dependent value for IPR, analogous to the early computations of delocalization in a specific eigenbasis in the atomic context. Averaging over several such slice-dependent values yields the mean IPR, which we denote by $\xi_{\text{IPR}}$. As in the case of the example of the disordered spin chain mentioned above, entanglement indicators can now be defined based on IPR, the new feature being that it is now adapted to CV systems. It is to be noted, though, that since IPR only estimates delocalization, it is in general nonzero even for separable states. Consequently, an entanglement indicator based on IPR does not vanish for unentangled states. Despite this feature, this indicator, as also other TEIs, have been found to track entanglement dynamics effectively in a variety of systems (Sharmila et al., 2019).

It is worth pointing out that this is only an instance of how the tomographic approach is useful in calculating both slice-dependent values and averaged values of all TEIs, and not merely those corresponding to IPR. This advantage is not available in the computation of entanglement measures such as SVNE which are obtained as a single quantity from the reconstructed density matrix. Slice-dependent indicators are hence the natural choice in comparing entanglement trends with EUR trends, as the latter are defined only for specific slices—either canonically conjugate slices, or arbitrarily chosen slices defined by noncommuting operators. Hence, the SVNE is not always expected to capture trends in EURs as well as the TEIs. In the later sections we have examined the role of SVNE versus the TEIs in this context.

We now proceed to define the TEIs. The slice-dependent entanglement indicator based on IPR is given by,

$$\epsilon_{\text{IPR}}(\theta_A, \theta_B) = 1 - \left[\eta_A(\theta_A) + \eta_B(\theta_B) - \eta_{AB}(\theta_A, \theta_B)\right], \tag{9}$$

where

$$\eta_{AB}(\theta_A, \theta_B) = \int_{-\infty}^{\infty} dX_{\theta_A} \int_{-\infty}^{\infty} dX_{\theta_B} \left[w(X_{\theta_A}, \theta_A; X_{\theta_B}, \theta_B)\right]^2 \tag{10}$$

is the two-mode IPR, and

$$\eta_k(\theta_k) = \int_{-\infty}^{\infty} dX_{\theta_k} \left[w_k(X_{\theta_k}, \theta_k)\right]^2 \ (k = A, B) \tag{11}$$

is the reduced subsystem IPR. Another useful TEI is given by

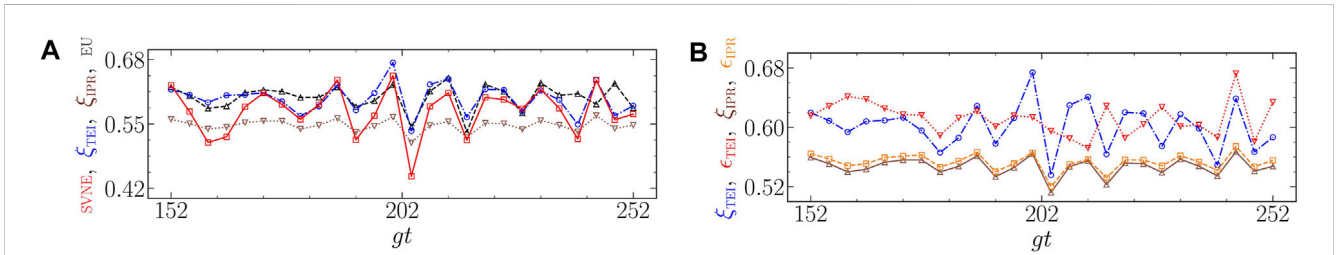

**FIGURE 1**
**(A)** Bipartite field-atom model. Initial state $|10; 0\rangle$ and weak nonlinearity ($\gamma/g = 0.01$). Entanglement indicators (red, blue, brown) and entropic sum uncertainty (EU) in the **x** and **p** quadratures (black) vs. scaled time. $0.2\times$ SVNE (red), $\xi_{\text{TEI}}$ (blue), $2\times[\xi_{\text{IPR}}-0.507]$ (brown), and $0.4\times[h(\mathbf{x})+h(\mathbf{p})-6.168]$ (black) *vs. gt*. Here **x** and **p** correspond to the $(0, 0)$ and $(\pi/2, \pi/2)$ tomographic slices, respectively. The ordinates have been appropriately scaled to enable ready comparison. At $t = 0$, $\xi_{\text{IPR}} = 0.507$ and EU = 6.168. **(B)** Bipartite field-atom model. Initial state $|10; 0\rangle$ and weak nonlinearity ($\gamma/g = 0.01$). Averaged tomographic entanglement indicators (blue, brown) and the corresponding tomographic entanglement indicators for the $(0, 0)$ slice (red, orange) vs. scaled time. $\xi_{\text{TEI}}$ (blue) and $0.8\times\epsilon_{\text{TEI}}(0, 0)$ (red), $2\times[\xi_{\text{IPR}}-0.507]$ (brown), and $2\times[\epsilon_{\text{IPR}}(0, 0)-0.507]$ (orange) *vs. gt*. The ordinates have been appropriately scaled to enable ready comparison. At $t = 0$, $\epsilon_{\text{IPR}}(0, 0) = 0.507$.

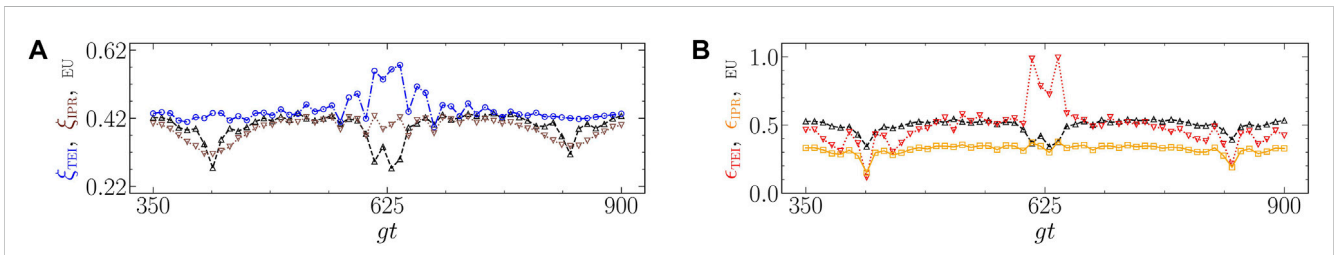

**FIGURE 2**
**(A)** Bipartite field-atom model. Initial state $|\alpha; 0\rangle$ ($|\alpha|^2 = 5$) and weak nonlinearity ($\gamma/g = 0.01$). Tomographic entanglement indicators (blue, brown) and entropic sum uncertainty (EU) in the **x** and **p** quadratures (black) vs. scaled time. $\xi_{\text{TEI}}$ (blue), $1.2\times[\xi_{\text{IPR}}-0.361]$ (brown), and $0.16\times[h(\mathbf{x})+h(\mathbf{p})-4.289]$ (black) *vs. gt*. The ordinates have been appropriately scaled to enable ready comparison. At $t = 0$, $\xi_{\text{IPR}} = 0.361$ and EU = 4.289. **(B)** Bipartite field-atom model. Initial state $|\alpha; 0\rangle$ ($|\alpha|^2 = 5$) and weak nonlinearity ($\gamma/g = 0.01$). Sum of the tomographic entanglement indicators in the **x** and **p** quadratures (red, orange), and entropic sum uncertainty (EU) in the same quadratures (black) vs. scaled time. $0.5\times[\epsilon_{\text{TEI}}(0, 0) + \epsilon_{\text{TEI}}(\pi/2, \pi/2)]$ (red), $0.5\times[\epsilon_{\text{IPR}}(0, 0) + \epsilon_{\text{IPR}}(\pi/2, \pi/2)-0.722]$ (orange), and $0.2\times[h(\mathbf{x})+h(\mathbf{p})-4.289]$ (black) *vs. gt*. The ordinates have been appropriately scaled to enable ready comparison. At $t = 0$, $\epsilon_{\text{IPR}}(0, 0) + \epsilon_{\text{IPR}}(\pi/2, \pi/2) = 0.722$ and EU = 4.289.

$$\epsilon_{\text{TEI}}(\theta_A, \theta_B) = S_A(\theta_A) + S_B(\theta_B) - S_{AB}(\theta_A, \theta_B) \quad (12)$$

where

$$S_k(\theta_k) = -\int_{-\infty}^{\infty} dX_{\theta_k} w_k(X_{\theta_k}, \theta_k) \log w_k(X_{\theta_k}, \theta_k) \quad (13)$$

is the subsystem tomographic entropy corresponding to subsystem $k$, and

$$S_{AB}(\theta_A, \theta_B) = -\int_{-\infty}^{\infty} dX_{\theta_A} \int_{-\infty}^{\infty} dX_{\theta_B} w_{AB}(X_{\theta_A}, \theta_A; X_{\theta_B}, \theta_B)$$
$$\times \log w_{AB}(X_{\theta_A}, \theta_B; X_{\theta_B}, \theta_B) \quad (14)$$

is the bipartite tomographic entropy (Sharmila et al., 2020; Sharmila et al., 2022). Note the similarity in form of Eq. 12 and quantum mutual information (QMI). In order to remove the dependence on $(\theta_A, \theta_B)$ from Eqs 9, 12, one averages over the quorum (comprising a set of indicators, each corresponding to different slices) to get

$$\xi_{\text{IPR}} = \langle \epsilon_{\text{IPR}}(\theta_A, \theta_B) \rangle \quad (15)$$

and

$$\xi_{\text{TEI}} = \langle \epsilon_{\text{TEI}}(\theta_A, \theta_B) \rangle. \quad (16)$$

In principle, one needs an infinite number of basis sets in the range $0 \leq \theta_A < \pi$ and $0 \leq \theta_B < \pi$, to obtain an entanglement indicator which compares with standard entanglement measures. An important aspect that needs to be addressed in this context, therefore, is the estimation of the optimal number of slices required in practice for a TEI to be useful. While this is sensitive to the specific system under consideration, we have observed that, for a variety of bipartite CV and HQ systems where a total number of quanta is conserved, averaging over approximately 25 slices equispaced between $[0, \pi)$ for both $\theta_A$ and $\theta_B$ suffices to capture the gross features.

The variance based uncertainties, entropic uncertainties and the corresponding bounds can be computed in a straightforward manner from tomograms. A useful and readily applicable procedure to compute the variance and all moments of quadrature observables (corresponding to a given tomographic slice) is given in Wünsche (1996). Our emphasis is on the computation of the tomographic information entropies that are directly relevant for EURs.

For any bipartite system the uncertainty relation of direct relevance to us pertains to canonically conjugate quadratures with variables $(x_1, p_1)$ and $(x_2, p_2)$ for the two subsystems. This EUR is given by

$$h(\mathbf{x}) + h(\mathbf{p}) \geqslant 2(1 + \ln \pi), \quad (17)$$

where $\mathbf{x} = (x_1, x_2)$, $\mathbf{p} = (p_1, p_2)$ and

$$h(\mathbf{x}) = -\int dx_1 dx_2 \, \varrho(x_1, x_2) \ln \varrho(x_1, x_2) \quad (18)$$

is the Shannon differential entropy [see, for instance, Hertz and Cerf (2019)]. Here, $\varrho(\mathbf{x})$ and $\varrho(\mathbf{p})$ are the probability distributions along the $\mathbf{x}$ and $\mathbf{p}$ quadratures respectively. Hence in the tomographic approach, $\varrho(x_1, x_2)$ is simply $w(X_{(\theta_A=0)}, \theta_A = 0; X_{(\theta_B=0)}, \theta_B = 0)$. The probability in the momentum quadrature is simply the two-mode tomographic slice obtained by setting $\theta_A$ and $\theta_B$ equal to $\pi/2$. A state is said to be *informationally squeezed* in $\mathbf{x}$ if $h(\mathbf{x}) < 1+ \ln \pi$. An analogous definition holds for the $\mathbf{p}$ quadrature. It is evident that the EUR can be now defined for different pairs of canonically conjugate slices. We may also consider the generalized EUR corresponding to the Rényi $q$-entropy in the place of the Shannon entropy in Eq. 18. In that case, an increase in $q$ leads to a decrease in the lower bound of EUR.

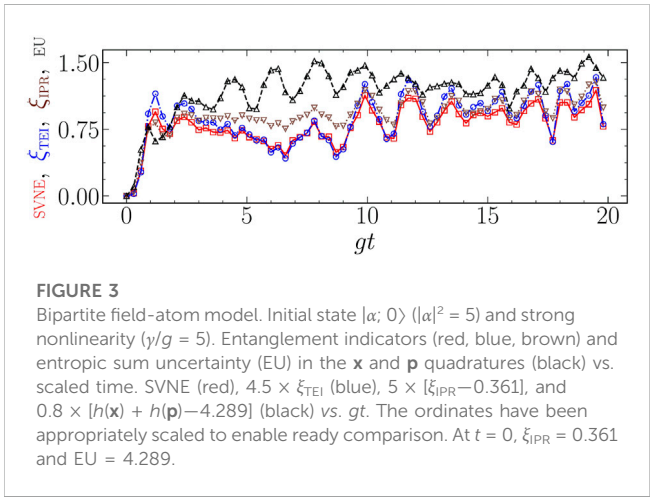

**FIGURE 3**
Bipartite field-atom model. Initial state $|\alpha; 0\rangle$ ($|\alpha|^2 = 5$) and strong nonlinearity ($\gamma/g = 5$). Entanglement indicators (red, blue, brown) and entropic sum uncertainty (EU) in the $\mathbf{x}$ and $\mathbf{p}$ quadratures (black) vs. scaled time. SVNE (red), $4.5 \times \xi_{\text{TEI}}$ (blue), $5 \times [\xi_{\text{IPR}} - 0.361]$, and $0.8 \times [h(\mathbf{x}) + h(\mathbf{p}) - 4.289]$ (black) vs. $gt$. The ordinates have been appropriately scaled to enable ready comparison. At $t = 0$, $\xi_{\text{IPR}} = 0.361$ and EU = 4.289.

In the next section we examine the reliability of bipartite TEIs in capturing the gross features of EURs in two generic systems, that are bipartite and tripartite respectively.

# 3 Results and discussions

## 3.1 Bipartite atom-field interaction model

The Hamiltonian for a radiation field with photon creation and annihilation operators $\hat{a}^\dagger$ and $\hat{a}$, propagating through a nonlinear atomic medium modelled as a multilevel oscillator with ladder operators $\hat{b}^\dagger$ and $\hat{b}$, is given by

$$\hat{\mathcal{H}}_1 = \omega \hat{a}^\dagger \hat{a} + \omega_0 \hat{b}^\dagger \hat{b} + \gamma \hat{b}^{\dagger 2} \hat{b}^2 + g\left(\hat{a}^\dagger \hat{b} + \hat{a} \hat{b}^\dagger\right). \quad (19)$$

Here, and in the rest of this paper, we set $\hbar = 1$. The field and atomic frequencies are respectively $\omega$ and $\omega_0$, $\gamma$ is the strength of the Kerr nonlinearity, and $g$ is the interaction strength. Since $\hat{\mathcal{N}} = \hat{a}^\dagger \hat{a} + \hat{b}^\dagger \hat{b}$ is a constant of the motion, it is convenient to examine the dynamics of both the atom and the field in the basis $|N - m\rangle_f \otimes |m\rangle_a$ denoted by

$|N - m; m\rangle$. Here, $m = 0, 1, 2, \ldots$ labels the atomic levels, and the suffixes $f$ and $a$ denote the field and atom respectively. The eigenvalues of $\hat{\mathcal{N}}$ are $N = 0, 1, 2, \ldots$. Consider the subsequent dynamics of an initial factored product of a photon number state or a coherent state, with the atom in an energy eigenstate. This can be understood to a large extent through analytical calculations carried out using realistic approximations, in the case of both weak and strong nonlinearity compared to the interaction strength. Further, in the former case, near-revivals have been predicted at specific instants of the scaled time $gt$ (Agarwal and Puri, 1989). The entanglement dynamics has been numerically investigated for these initial states, and quantified in terms of SVNE and SLE of the field state in (Sudheesh et al., 2006). It has been reported that, for weak nonlinearity and an initial Fock state, the entanglement dips at instants of both approximate revivals and fractional revivals.

We have numerically examined how efficiently the TEIs capture these aspects of the dynamics. We illustrate these features for the atom in the ground state, $\omega = \omega_0 = 1$ and the field initially either in a Fock state or a CS $|\alpha\rangle$, both for weak ($\gamma/g \ll 1$) and strong ($\gamma/g \gg 1$) nonlinearity. We have also investigated the temporal dynamics of EUR corresponding to the pure bipartite state. We examine the extent to which the TEIs mimic the dynamics of SVNE and also identify the entanglement indicators which mimic the dynamics of EUR. Our results are briefly summarized below.

We first consider the field initially in the Fock state $|N = 10\rangle$. For weak nonlinearity, it is straightforward to see that for $\theta_A = \theta_B = \theta$, both the bipartite and the subsystem tomograms are independent of $\theta$. This follows from the conservation of $\hat{\mathcal{N}}$ and the fact that the field is initially in a photon number state. The TEIs extremize at points where SVNE and EUR also extremize. However, as illustrated in Figure 1A, while the general trends agree, the behavior is very sensitive to the time interval considered. As an example, for $gt$ between 240 and 248, the TEIs and SVNE do not mirror EUR. Similar conclusions follow when the tomographic slices ($\pi/4$, $\pi/4$) and ($3\pi/4$, $3\pi/4$) are compared, instead of (0, 0) and ($\pi/2$, $\pi/2$) as in the figure. Further, while $\epsilon_{\text{TEI}}$ and $\xi_{\text{TEI}}$ do not mimic each other, $\epsilon_{\text{IPR}}$ and $\xi_{\text{IPR}}$ are relatively more similar in their trends (see Figure 1B). This is evidence that IPR is dominated by contributions for which $\theta_A = \theta_B$, while TEI has significant contributions from regions in which $\theta_A \neq \theta_B$. This feature is more pronounced for large $N$. The subsystem EUR dynamics is not captured by TEIs or SVNE. We have also verified that there is no entropic squeezing for weak nonlinearity.

For strong nonlinearity, the atom effectively behaves like a two-level system, with periodic exchange of energy with the field (Agarwal and Puri, 1989). We have verified that SVNE, TEIs and EUR emulate each other. Further, the EUR corresponding to the field subsystem is also captured by the TEIs. However, for large $N$, the changes in $\xi_{\text{IPR}}$ with scaled time are small in comparison with the corresponding changes in $\xi_{\text{TEI}}$. In this sense, $\xi_{\text{TEI}}$ reflects the EUR dynamics better than $\xi_{\text{IPR}}$. As in the case of weak nonlinearity, $\xi_{\text{IPR}}$ and $\epsilon_{\text{IPR}}$ are similar in their trends, while $\xi_{\text{TEI}}$ and $\epsilon_{\text{TEI}}$ are dissimilar.

The EUR based on the Rényi entropy is alike in dynamics to the EUR corresponding to the Shannon entropy, independent of the extent of nonlinearity. Therefore, in what follows, we shall only examine the EUR based on the Shannon entropy.

Next, we consider the field to be initially in a CS $|\alpha\rangle$ ($|\alpha|^2 = 5$). In this case the TEIs depend on the specific slice considered, in contrast to the preceding case. For the **x** and **p** quadratures and weak nonlinearity, as in Figure 2A, the dynamics of EUR is not reflected in that of the TEIs for all time intervals (as before), although $\xi_{IPR}$ performs better than $\xi_{TEI}$, on the average. Further, the dynamics of $\epsilon_{IPR}$ is similar to $\xi_{IPR}$, and, overall, it captures the trends of EUR better than $\epsilon_{TEI}$ (Figure 2B). For strong nonlinearity, the main results are strikingly different from the case of an initial Fock state reported earlier (see Figure 3). SVNE and TEIs have similar behaviour in dynamics. However, EUR is not captured by either, except during short time intervals (e.g., *gt* between 0 and 1 and also between 15 and 20). The $\epsilon$ indicators mimic the corresponding $\xi$ indicators reasonably well.

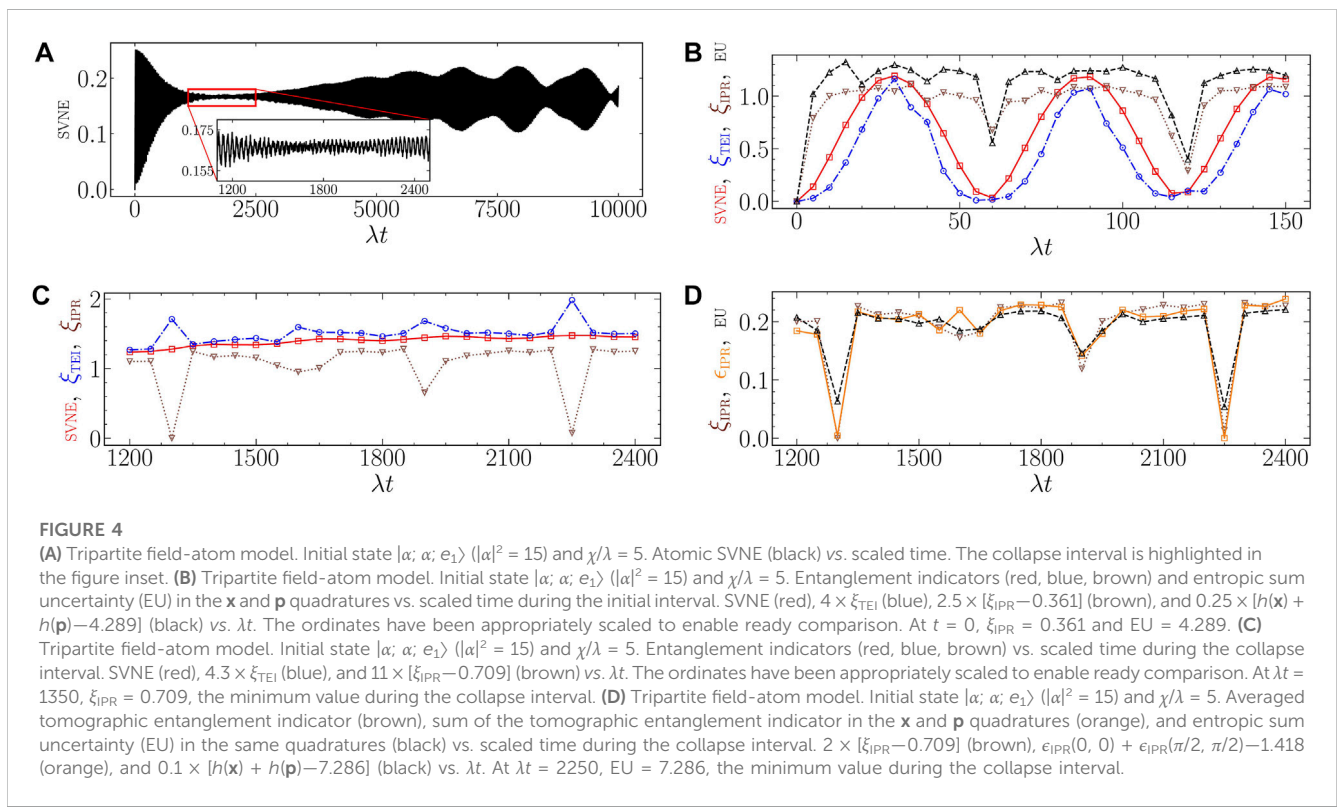

**FIGURE 4**
**(A)** Tripartite field-atom model. Initial state $|\alpha; \alpha; e_1\rangle$ ($|\alpha|^2 = 15$) and $\chi/\lambda = 5$. Atomic SVNE (black) *vs.* scaled time. The collapse interval is highlighted in the figure inset. **(B)** Tripartite field-atom model. Initial state $|\alpha; \alpha; e_1\rangle$ ($|\alpha|^2 = 15$) and $\chi/\lambda = 5$. Entanglement indicators (red, blue, brown) and entropic sum uncertainty (EU) in the **x** and **p** quadratures vs. scaled time during the initial interval. SVNE (red), $4 \times \xi_{TEI}$ (blue), $2.5 \times [\xi_{IPR} - 0.361]$ (brown), and $0.25 \times [h(\mathbf{x}) + h(\mathbf{p}) - 4.289]$ (black) *vs.* $\lambda t$. The ordinates have been appropriately scaled to enable ready comparison. At $t = 0$, $\xi_{IPR} = 0.361$ and EU = 4.289. **(C)** Tripartite field-atom model. Initial state $|\alpha; \alpha; e_1\rangle$ ($|\alpha|^2 = 15$) and $\chi/\lambda = 5$. Entanglement indicators (red, blue, brown) vs. scaled time during the collapse interval. SVNE (red), $4.3 \times \xi_{TEI}$ (blue), and $11 \times [\xi_{IPR} - 0.709]$ (brown) *vs.* $\lambda t$. The ordinates have been appropriately scaled to enable ready comparison. At $\lambda t = 1350$, $\xi_{IPR} = 0.709$, the minimum value during the collapse interval. **(D)** Tripartite field-atom model. Initial state $|\alpha; \alpha; e_1\rangle$ ($|\alpha|^2 = 15$) and $\chi/\lambda = 5$. Averaged tomographic entanglement indicator (brown), sum of the tomographic entanglement indicator in the **x** and **p** quadratures (orange), and entropic sum uncertainty (EU) in the same quadratures (black) vs. scaled time during the collapse interval. $2 \times [\xi_{IPR} - 0.709]$ (brown), $\epsilon_{IPR}(0, 0) + \epsilon_{IPR}(\pi/2, \pi/2) - 1.418$ (orange), and $0.1 \times [h(\mathbf{x}) + h(\mathbf{p}) - 7.286]$ (black) *vs.* $\lambda t$. At $\lambda t = 2250$, EU = 7.286, the minimum value during the collapse interval.

From the above results we conclude that for weak nonlinearity and a pure bipartite state, the efficacy of both the TEIs in mimicking the dynamics of EUR is not very sensitive to the precise nature of the initial field state. Further, both $\xi_{IPR}$ and $\xi_{TEI}$ are comparable in their performance. In what follows, therefore, we will only consider strong nonlinearity and bipartite entanglement in a mixed state. In particular, we will compare how well the dynamics of SVNE, $\xi_{IPR}$ and $\xi_{TEI}$ follow that of EUR.

Investigation of the effect of a *finite* number of atomic levels (in contrast to the model considered in this section) in understanding the quantum-classical divide is an aspect of considerable interest. In the classical context consider a coarse-grained phase space. Analysis of the Poincaré recurrences of the dynamical variables to cells in this space augmented with time series analysis of the variables considered reveal that, if the first return time distribution to a cell in phase space is spiky, the Lyapunov exponent obtained through the time series analysis is zero, implying non-chaotic dynamics. However, corresponding to exponential first return time distributions, the Lyapunov exponent is positive, signaling chaotic dynamics (Balakrishnan et al., 2001; Balakrishnan and Theunissen, 2001). In the quantum context, treating the observable as a dynamical variable in an appropriate coarse-grained phase space, similar studies on their ergodicity properties have been undertaken. The results show that if the number of atomic levels is infinite, as in the model above, the classical relation between the return time distribution and the Lyapunov exponent continues to hold (Sudheesh et al., 2009). However, the case of a 3-level atom interacting with one or two radiation fields, with the mean photon number treated as a dynamical variable, exhibits surprising results (Shankar et al., 2014). In specific situations, spiky first return distributions accompany positive Lyapunov exponents, while exponential distributions could go hand in hand with vanishing Lyapunov exponents. In the next section, we therefore consider the dynamics of the mean photon number in a tripartite model with a finite number of atomic levels in the context of EURs and entanglement, in order to examine whether a reduction of the atomic levels to a finite number affects the inferences drawn in this section.

## 3.2 The tripartite HQ model

We consider a tripartite model of a Λ-atom with energy levels $\{|e_1\rangle, |e_2\rangle, |e_3\rangle\}$, interacting with two radiation fields $F_i$, with photon creation and annihilation operators $\hat{a}_i^\dagger$ and $\hat{a}_i$, and frequency $\Omega_i$ respectively ($i = 1, 2$). $F_i$ mediates the $|e_i\rangle \leftrightarrow |e_3\rangle$ transition, where $|e_3\rangle$ is the highest energy state. The $|e_1\rangle \leftrightarrow |e_2\rangle$ transitions are dipole forbidden. The tripartite Hamiltonian wth zero detuning is

$$\hat{\mathcal{H}}_2 = \sum_{k=1}^{3} \omega_k \hat{\sigma}_{kk} + \sum_{i=1}^{2} \left[ \Omega_i \hat{a}_i^\dagger \hat{a}_i + \chi \hat{a}_i^{\dagger 2} \hat{a}_i^2 + \lambda f\left(\hat{a}_i^\dagger \hat{a}_i\right)\left(\hat{a}_i \hat{\sigma}_{3i} + \text{h.c.}\right) \right].$$

(20)

Here, $\hat{\sigma}_{jk} = |e_j\rangle\langle e_k|$ $(j, k = 1, 2, 3)$ are the atomic ladder operators, $\{\omega_k\}$ are positive constants, $\chi$ is the strength of the field Kerr nonlinearity, $\lambda$ is the interaction strength and $f\left(\hat{a}_i^\dagger \hat{a}_i\right)$ is the intensity-dependent field-atom coupling. Different forms of $f$ have been examined in the literature. An interesting case corresponds to $f = [1 + \kappa \left(\hat{a}_i^\dagger \hat{a}_i\right)]$, where $\kappa$ is a tunable parameter, with an unentangled initial state comprising a factored product of two CS $|\alpha\rangle$ and the atom in $|e_1\rangle$. The subsequent temporal dynamics corresponding to this case has been investigated extensively. A spectacular bifurcation cascade has been reported as $\kappa$ is fine-tuned between 0 and 1 (Laha et al., 2016). We set $\kappa = 0$ in our investigations, and examine the dynamics in terms of the scaled time $\lambda t$. Two time scales of direct relevance are characterized approximately by the initial interval (0, 150), and the collapse region (1200, 2400) of $\lambda t$ (Figure 4A). In the latter interval, SVNE collapses to an approximately constant value. This collapse is captured in the dynamics of the mean photon number corresponding to the field considered. For illustrative purposes, and without loss of generality, we set $|\alpha|^2 = 15$ and $\chi/\lambda = 5$ in our numerical computations. Although the tripartite system is described by a pure state, the bipartite field subsystem (obtained by tracing out the atomic subsystem) is mixed and is therefore ideally suited for examining mixed state bipartite entanglement. We have verified that the inferences obtained by considering the two different pairs of canonically conjugate slices mentioned earlier, are similar.

We now summarize the main results. Consider, first, the initial time interval. The broad features displayed by either field subsystem are similar (though not identical). This is because the atomic subsystem in this case is only weakly entangled with the field subsystems. Hence we have carried out the computation of SVNE with the density matrix corresponding to one of the fields. From Figure 4B it is evident that the entanglement dips in SVNE at $\lambda t = 60$ and 120, for instance, are also reflected in $\xi_{\text{TEI}}, \xi_{\text{IPR}}$ and EUR. However, EUR and $\xi_{\text{IPR}}$ resemble each other more closely in the interval considered, whereas SVNE and $\xi_{\text{TEI}}$ show similar roughly oscillatory behavior. We have verified that the trends in the $\epsilon$ indicators are similar to those of the corresponding $\xi$ indicators.

We now consider the dynamics during the time interval of collapse. In stark contrast to the foregoing observations, the dynamics of SVNE is not mirrored in any of the TEIs. Whereas SVNE collapses to a nearly constant value over the entire interval, the TEIs extremize in the neighbourhood of $\lambda t = 1300, 1900$ and 2250 (see Figure 4C). Further, it is evident from Figure 4D, that the dynamics of $\xi_{\text{IPR}}$ and that of EUR are remarkably similar. This is also reflected in the dynamics of $[\epsilon_{\text{IPR}}(0, 0) + \epsilon_{\text{IPR}}(\pi/2, \pi/2)]$.

## 4 Conclusion

We have examined the dynamics of bipartite entanglement of both pure and mixed CV states in two generic models of atom-field interaction. A primary purpose of this investigation has been to compare the manner in which TEIs on the one hand, and SVNE on the other, mimic the dynamics of EURs. Further, we have identified the TEI which closely tracks the temporal trends of EURs under different situations, such as, weak *versus* strong nonlinearity, and mixed *versus* pure states. The importance of this approach is emphasized by the fact that SVNE depends on the density operator, and it is in general a formidable task to reconstruct the density operator from tomograms in generic CV and large HQ systems. Our findings from the numerical simulations using the tomographic approach employed in this paper provide a viable alternative to assess bipartite entanglement in such cases. The broad picture that emerges from the present work is that SVNE and $\xi_{\text{TEI}}$ resemble each other in their gross features, but do not follow the trends in EURs in general. An interesting outcome of our investigation is that the efficacy of $\xi_{\text{IPR}}$ in mimicking EUR is very reasonable. This is an illustration of the deficiency of an entanglement *measure* in reflecting the dynamics of EUR, and of the usefulness of slice-dependent indicators.

## Data availability statement

The original contributions presented in the study are included in the article/Supplementary Material, further inquiries can be directed to the corresponding author.

## Author contributions

SP carried out the numerical computations, generated the figures and wrote the first draft of the manuscript. SL, VB, and SR contributed to the conception and design of the study, and also revised and produced the final manuscript.

## Funding

This work was supported in part by a grant from Mphasis to the Center for Quantum Information, Communication and Computing (CQuICC).

## Conflict of interest

The authors declare that the research was conducted in the absence of any commercial or financial relationships that could be construed as a potential conflict of interest.

## Publisher's note

All claims expressed in this article are solely those of the authors and do not necessarily represent those of their affiliated organizations, or those of the publisher, the editors and the reviewers. Any product that may be evaluated in this article, or claim that may be made by its manufacturer, is not guaranteed or endorsed by the publisher.